\documentclass[showpacs,twocolumn,superscriptaddress,prb,preprintnumbers,nofootinbib]{revtex4}
\usepackage{amsmath,amssymb,bm,graphicx}

\begin{document}

\title{Strong quantum effects in an almost classical antiferromagnet on a kagome lattice}

\author{A. L. Chernyshev}
\affiliation{Department of Physics and Astronomy, University of California, Irvine, California
92697, USA}
\date{\today}
\begin{abstract}  
Two ubiquitous features of frustrated spin systems stand out:  massive degeneracy of their ground states and flat, or 
dispersionless, excitation branches. In real materials, the former is frequently lifted by secondary interactions or quantum 
fluctuations, in favor of an ordered or spin-liquid state, 
but the latter often survive. 
We demonstrate that  flat modes may precipitate 
remarkably strong quantum effects even in the systems that are otherwise written off as almost entirely classical.
The resultant spectral features should be reminiscent of the quasiparticle breakdown in 
quantum systems, only here the effect is strongly amplified by the flatness of  spin-excitation branches,
leading to the damping that is not vanishingly small even at $S\!\gg\!1$. 
We  provide a  theoretical analysis of   excitation spectrum 
of the   $S=5/2$ iron-jarosite to illustrate our findings and to 
suggest  further studies of this and other frustrated spin systems. 
\end{abstract}
\pacs{75.10.Jm, 	
      75.30.Ds,     
      75.50.Ee, 	
      75.40.Gb,     
      78.70.Nx     
}
\maketitle

\section{Introduction}

Ever since their inception in the 1950s,\cite{Wannier,Villain} frustrated spin systems have been a
source of new ideas for a wide variety of problems;  
unconventional superconductivity,\cite{Anderson} order-by-disorder phenomena, \cite{Villain80} and
correlated spin-liquid states,\cite{Balents10}  are among them.
In the core of this fertility is the near degeneracy between a vast number of spin configurations, 
originating from   competing interactions that are favoring mutually exclusive ground states.   
In frustrated magnetic materials and   their models this massive   
degeneracy is responsible 
for an extreme sensitivity to subtle symmetry breaking effects,\cite{Huse92,Chalker92} strongly amplified role of 
subleading coupling terms,\cite{Harris92} hierarchy of emergent energy scales,\cite{Taillefumier14} 
and order-by-disorder effects by 
thermal,\cite{Reimers93} and quantum fluctuations.\cite{Chubukov92,ChZh14,Jackeli15} 

\begin{figure}[b]
\vskip -0.5cm
\includegraphics[width=0.99\columnwidth]{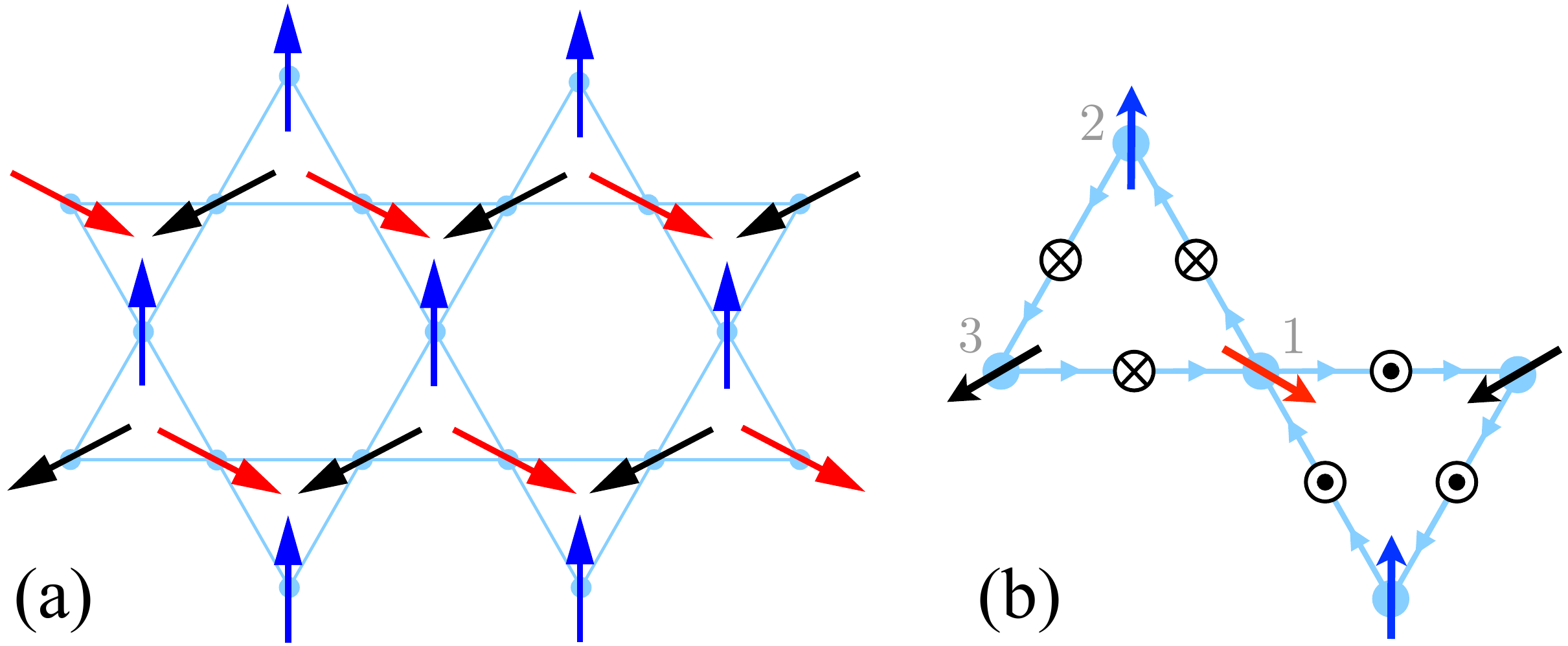}
\caption{(Color online)\ \  (a) 
${\bf q}=0$ type of spin ordering on the kagome lattice. (b)  Directions of the DM vectors. 
Arrows on the bonds show the ordering of  ${\bf S}_i$ and ${\bf S}_j$  in \eqref{HDM}.
}
\label{Fig:DM}
\end{figure}

Concomitant  of the ground-state degeneracy is another hallmark feature of the
frustrated  spin systems: flat excitation branches at
low energies.\cite{Chalker92,Harris92,Chubukov92,Matan06,Yildirim06,Matan14,Petrenko}  
They owe their origin to both   the topological structure of the underlying lattices that facilitate frustration and
 the insufficient constraint on the manifold of spin configurations.
A subclass of frustrated magnets that exhibits  flat modes prominently 
is the kagome-lattice  antiferromagnets.
\cite{Matan10,Matan14,Helton07,Lee12,Yan11,Balents10,Mambrini00,Iqbal13,Rousochatzakis14}
 Under the influence of  subleading interactions,
majority of the known kagome-lattice  antiferromagnets order magnetically  with  spins forming
non-collinear structures\cite{Zorko13,Yoshida12,Matan06,Yildirim06,Elhajal02} 
that are often reminiscent of the classical $120^\circ$ motif on 
each triangle, Fig.~\ref{Fig:DM}(a). 
Such a pattern is also emblematic of the geometric frustration, manifesting
a compromise reached by spins locally to partially satisfy their antiferromagnetic trends. 

The following  aspect  of this picture is crucial. 
The non-collinearity of the ordered spin pattern implies strong nonlinear, anharmonic effects.\cite{RMP13}
The role of  such effects in the ground-state selection of frustrated systems has been recognized 
since the early days of the field\cite{Chalker92,Chubukov92,Henley93} and, recently,  an accurate, systematic treatments of the 
quantum order-by-disorder effect due to them has received significant attention.\cite{ChZh14,Gotze15}

On the other hand, their role in the excitation spectra of the kagome-lattice antiferromagnets 
has been hardly touched upon. In this work, we demonstrate that the nonlinear terms can be particularly important in 
the spectral properties of the flat-band frustrated magnets, leading to spectacularly strong quantum effects even 
in the systems that are assumed almost classical. The resultant spectral features bear a remarkable similarity 
to the quasiparticle breakdown signatures in quantum spin-
and Bose-liquids, such as superfluid $^4$He,\cite{Zaliznyak,Zheludev}  
which exhibit characteristic termination points and  
ranges of energies where single-particle excitations are not well-defined and are dominated instead by 
broad continua.

It is usually assumed that such drastic effects can only occur in the systems that are inherently quantum
in nature.\cite{RMP13,Zaliznyak,Zheludev} In our case, their origin is in the 
near resonance decay  of the ``normal'', i.e. dispersive, modes into pairs of the flat-mode excitations 
facilitated by the nonlinear couplings. As such, the effect is strongly amplified by the density of states 
of the flat modes and is very significant even for large-spin systems that can  otherwise appear as purely classical,
resulting in the damping effect $\Gamma_{\bf k}/\varepsilon_{\bf k}\!\sim\!1$.
 While in the following we give a detailed account of the spectral properties of a specific kagome-lattice 
 antiferromagnet, $S=5/2$ Fe-jarosite,  encouraging its further investigation by 
 inelastic neutron scattering, the outlined scenario should be applicable to a wide variety of other flat-band 
 frustrated spin systems.\cite{Petrenko,Matan14,Balents_honeycomb}

The paper is organized as follows. In Section~\ref{Sec:H} we provide a general argument for the resonant-like
decay to exist in the frustrated flat-band systems and lay out a qualitative expectation, which highlights an unusual 
phenomenon: decays remain significant even in the $S\rightarrow\infty$ limit. In Section~\ref{Sec:FeJ} we provide our results 
for the decay-dominated spectral features in Fe-jarosite. Section~\ref{Sec:summ} gives a brief summary.
Technical details are delegated to   Appendix~\ref{app0}.

\section{Nonlinear coupling and resonant-like decays}
\label{Sec:H}

Because of the noncollinear structure of the ground-state spin configuration, the 
interacting spin excitations in the kagome-lattice antiferromagnets 
are described by 
\begin{eqnarray}
\hat{\cal H} = \sum_{{\bf k}\mu} \varepsilon_{\mu\bf k}b^\dag_{\mu{\bf k}} b^{\phantom\dag}_{\mu{\bf k}}
+\frac12\sum_{\bf p+q=k}\Phi^{\nu\eta\mu}_{\bf qp;k} 
b_{\nu\bf q}^\dagger b_{\eta\bf p}^\dagger  b_{\mu\bf k} + \textrm{h.c.} \,,
\label{Heff}
\end{eqnarray}
where the first term accounts for the  spin-wave energies 
while the second is an outcome of the anharmonic coupling of spins that results in the 
mutual transitions between excitation branches, see Appendix~\ref{app0}  for details. 
Specifically, it couples dispersive excitations 
with the flat modes, allowing for the resonance-like decay of the former into the pairs of the latter.
We note that this general form of bosonic Hamiltonian (\ref{Heff}) occurs in a variety of contexts,
including other frustrated antiferromagnets with noncollinear order,\cite{tri06,RMP13} 
as well as spin-liquids,\cite{Zh06} valence-bond solids,\cite{Vojta} and Bose-liquids.\cite{Zaliznyak}

The full extend of the 
the $1/S$-expansion also involves quartic and  source cubic terms, see Refs.~\onlinecite{RMP13,ChZh14}.
 Then, the magnon Green's function for the branch $\mu$ is 
\begin{eqnarray}
G_{\mu{\bf k}}^{-1}(\omega)= \omega-\varepsilon_{\mu{\bf k}}-\Sigma_{\mu{\bf k}}(\omega) \,,
\label{GF}
\end{eqnarray}
in which the self-energy $\Sigma_{\mu{\bf k}}(\omega)$ includes all such terms.
However, it is only decay terms in (\ref{Heff})   that are responsible for the resonance-like decay
phenomenon discussed in this work. Given the off-resonance character of the source term, the Hartree-Fock nature 
of the quartic terms, and the large-$S$ limit of the problem, one can safely approximate the self-energy by its 
on-shell imaginary part, i.e.  by
$\Sigma_{\mu{\bf k}}(\omega)|_{\omega=\varepsilon_{\mu{\bf k}}}\!\approx\!  -i\Gamma_{\mu{\bf k}}$.
The decay rate  $\Gamma_{\mu{\bf k}}$  in the  lowest-order  approximation is given by
\begin{eqnarray}
\Gamma_{\mu{\bf k}}=\frac{\pi}{2}\sum_{{\bf q},\nu\eta}
|\Phi^{\nu\eta\mu}_{{\bf q},{\bf k}-{\bf q};{\bf k}}|^2
\delta\left(\varepsilon_{\mu\bf k} - \varepsilon_{\nu\bf q} - \varepsilon_{\eta{\bf k}-{\bf q}}\right),
\label{Gamma}
\end{eqnarray}
where    sum is over the branches of the decay products and an explicit form of the vertex 
$\Phi^{\nu\eta\mu}_{{\bf q},{\bf k}-{\bf q};{\bf k}}$ is given in Appendix~\ref{app0}. 
With that, evaluation of the spectral function $A_{\mu{\bf k}}(\omega)\!=\!-(1/\pi){\rm Im}G_{\mu{\bf k}}(\omega)$ is
also straightforward.

The anharmonic cubic terms appear in the Hamiltonians of the noncollinear magnets via
bosonization of the terms that have a form $\sim S_i^zS_j^{x(y)}$ in the local reference frame of the ordered 
moments.\cite{RMP13} Because of that,   cubic vertices in  (\ref{Gamma})  necessarily scale with the spin value 
as $\Phi^{\nu\eta\mu}_{{\bf q},{\bf p};{\bf k}}\propto\!\sqrt{S}$. 
Since the energies of the decay products scale as $\varepsilon_{\nu\bf q} \propto S$, 
it follows that $\Gamma_{\mu{\bf k}}$ in (\ref{Gamma}) must be  spin-independent.
Contributions to the decay rate
from the higher-order terms should then follow a natural $1/S$ expansion 
with  the exception  of some special contours in ${\bf k}$-space where a $\log(S)$ 
enhancement in (\ref{Gamma}) is produced due to van Hove singularities of the two-magnon 
continuum.\cite{tri06,RMP13}
Therefore, one can conclude that for magnets with  $S \gg 1$, 
 damping of  higher-energy magnetic excitations  
due to decays into  lower-energy ones must be small compared to the excitation energy.
In other words,  generally, $\Gamma_{\mu{\bf k}}/\varepsilon_{\mu\bf k} \propto 1/S$ and
thus one expects that effects of   decays can be significant only for low-$S$, quantum magnets.\cite{RMP13}

Here we offer a general scenario in which this seemingly invincible logic fails dramatically.
If both decay products belong to the flat modes with  constant energy $\varepsilon_1$, 
a remarkably stronger effect must be taking place. 
Namely, in this case the self-energy of the dispersive modes exhibits an essential singularity at the energy  
$2\varepsilon_1$,  and, formally, the linewidth $\Gamma_{\mu{\bf k}}$ in (\ref{Gamma}) is infinite
at this energy, the effect we refer to as the resonance-like decay. 

In reality,  quantum fluctuations of the same origin, i.e. coming from the anharmonic 
cubic terms,  also generate effective further-neighbor $J_2$ spin 
couplings, \cite{Chubukov92,ChZh14} which necessarily warp the flat mode
and thus provide natural means  of regularizing this essential singularity.
However, the resultant fluctuation-induced bandwidth of the flat mode is now $S$-independent, 
$\delta\varepsilon_{\nu\bf q}\!\propto\!{\cal O}(S^0)$,
so that the regularized resonance-like broadening  in the vicinity of  $2\varepsilon_1$ must scale together with the 
excitation energy: $\Gamma_{\mu{\bf k}}\propto\varepsilon_{\mu\bf k}\propto S$. This qualitative 
consideration implies a spectacular quantum effect: a very strong damping, 
eliminating spectral weight  from the respective energy range even in large-$S$ magnets.
Thus, frustration provides necessary and sufficient ingredients for the proliferation
of intrinsically quantum phenomenon of decays into inherently classical spin systems.

Altogether, we predict that anomalous broadening and a wipe-out of the spectral 
weight, associated with the resonant-like decays, 
should be common in the spectra of the flat-band frustrated systems.
In practice, we argue that the quasiparticle breakdown with characteristic termination points and  
ranges of energies dominated by broad continua must be present in   $S\!=\!5/2$ 
kagome-lattice  Fe-jarosite.

\begin{figure*}
\includegraphics[width=0.99\columnwidth]{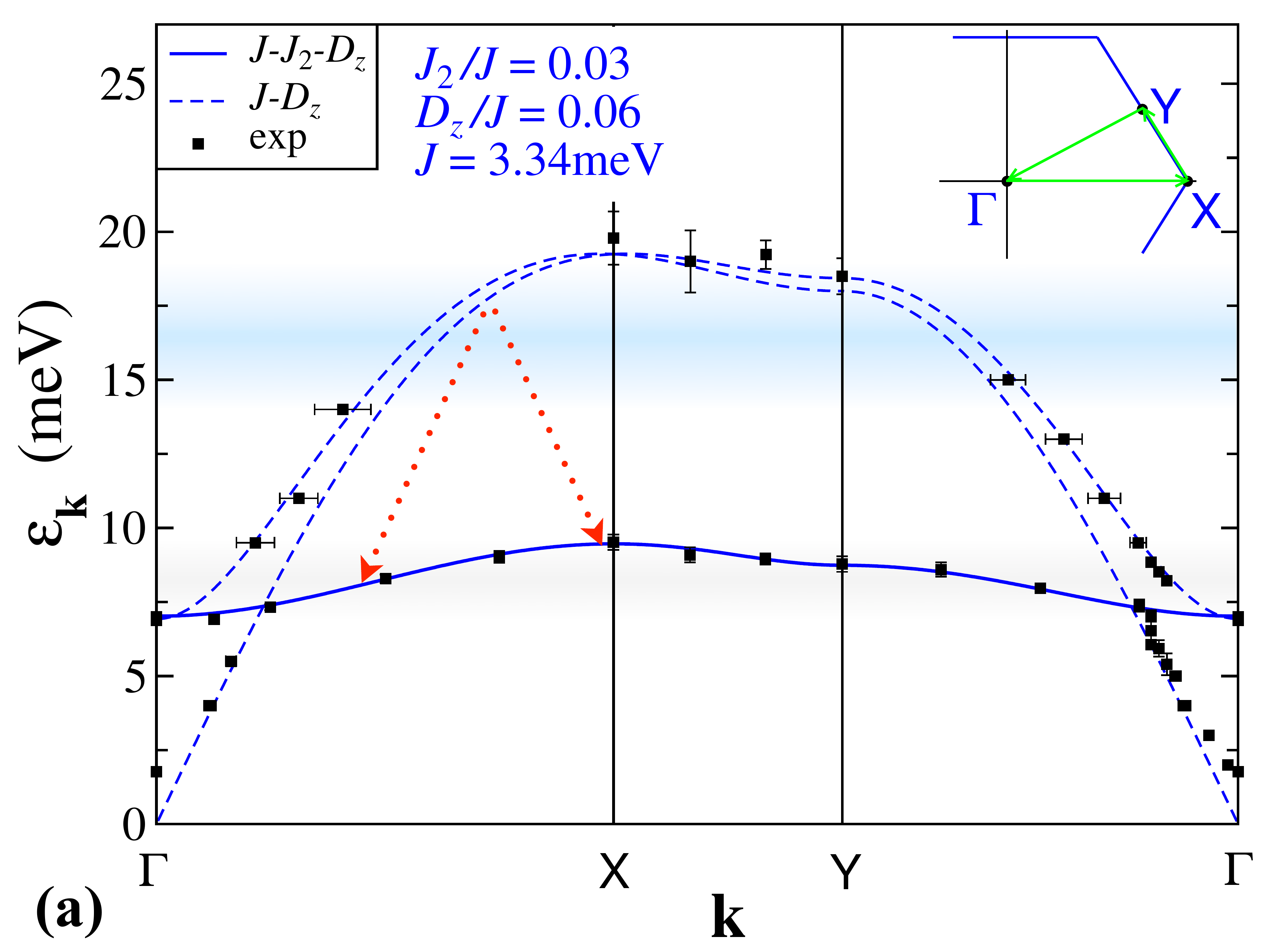} \quad\quad
\includegraphics[width=0.99\columnwidth]{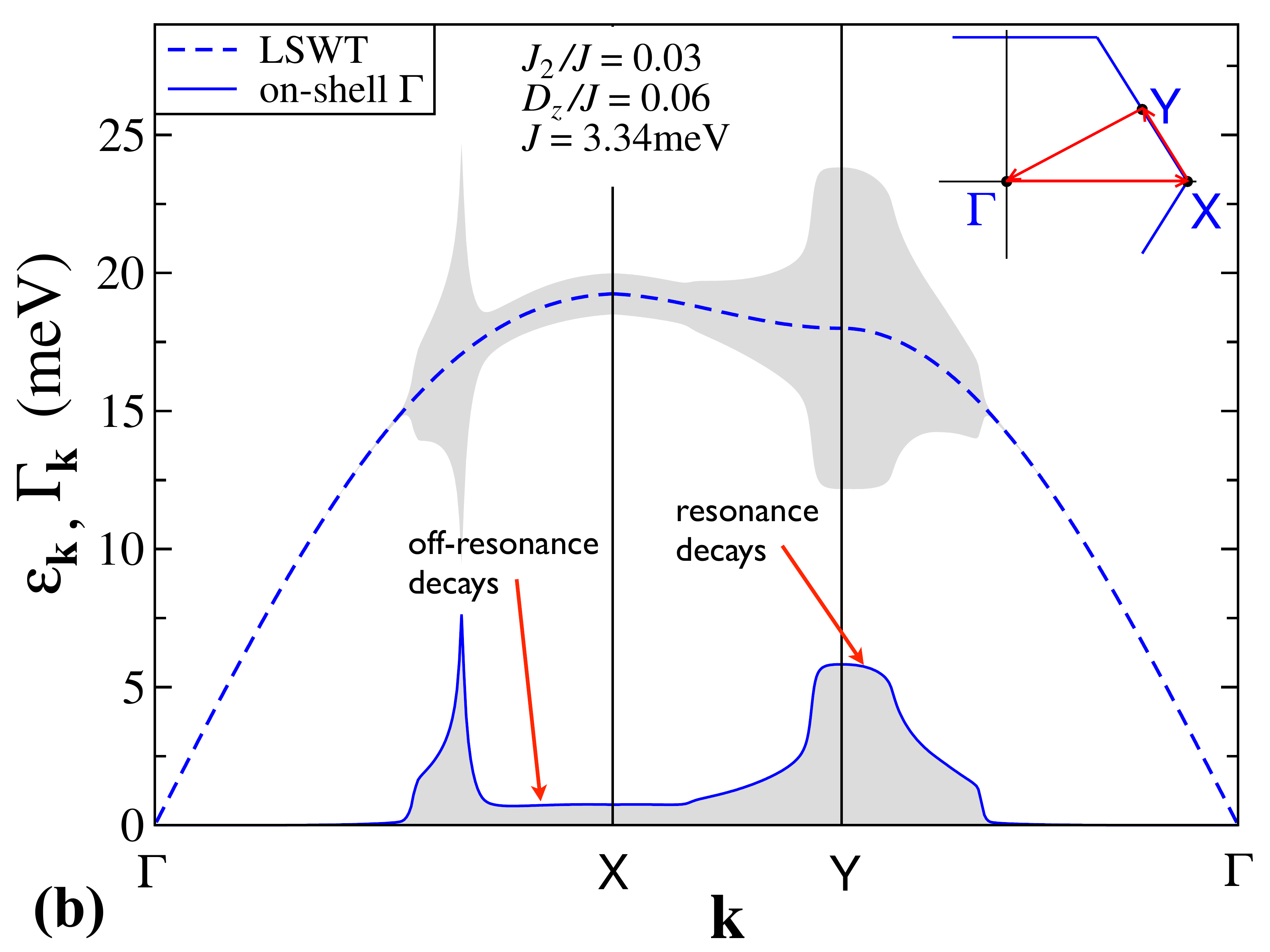}
\caption{(Color online)\ \
(a) Neutron-scattering data from \cite{Matan06}  
 along the $\Gamma$XY$\Gamma$ path (inset). Lines are linear spin-wave theory fits of the
dispersive modes (dashed) using (\ref{HDM}), and the flat mode (solid) with $J_2$ added to 
(\ref{HDM}), parameters are as shown. Lower shaded area highlights the flat band, 
upper is the set of energies of two flat modes. Arrows imply a decay process into two flat modes.
(b) Lower curve with the shading is the on-shell $\Gamma_{{\bf k}}$ from (\ref{Gamma}).
Dashed line is the linear spin-wave theory energy of the gapless dispersive mode from (a), 
shaded area shows the  half-width boundaries of a lorentzian peak,
$\varepsilon_{\bf k}\pm\Gamma_{{\bf k}}$.}
\label{Fig:wk_J2}
\end{figure*}

\section{Fe-jarosite}
\label{Sec:FeJ}

In realistic kagome-lattice antiferromagnets, the degeneracy within the manifold of 
classical $120^\circ$ states is, most commonly, lifted by the  
symmetry-breaking Dzyaloshinskii-Moriya (DM) terms, 
\cite{Zorko13,Elhajal02,Cepas08}
yielding  the  Hamiltonian that closely describes   Fe-jarosite \cite{Matan06,Yildirim06} and other systems
\cite{Zorko08,Yoshida12,francisite15}
\begin{equation}
\hat{\cal H} = \sum_{\langle ij\rangle}\big( J {\bf S}_i \cdot{\bf S}_j + {\bf D}\cdot {\bf S}_i \times{\bf S}_j\big)\ ,
\label{HDM}
\end{equation}
where summation is over the nearest-neighbor bonds and   
${\bf D}\!=\!(0,0,\mp D_z)$ on the  up/down triangles with 
the order of the site indices in  (\ref{HDM}) shown in Fig.~\ref{Fig:DM}(b). 
The out-of-plane DM interaction lifts the degeneracy and selects the  ${\bf q}\!=\!0$ ground state, i.e. 
a ``ferro''-120$^\circ$ pattern, Fig.~\ref{Fig:DM}(a). 
A small in-plane  DM term\cite{Yildirim06}  is neglected  for simplicity.

Given the large spin value, $S\!=\!5/2$, we estimate that the ordered moment should be nearly 90\% of its 
classical value.\cite{ChZh15} 
Similarly, the results of the earlier neutron scattering in  Fe-jarosite\cite{Matan06} 
have been interpreted as fully describable by the  linear spin-wave theory,\cite{Yildirim06} 
a construction whose validity we question next.

Our Fig.~\ref{Fig:wk_J2}(a) shows the linear spin-wave theory fits of the neutron-scattering data \cite{Matan06} 
using model (\ref{HDM}) where three distinct excitations branches are easy to identify. 
The DM anisotropy shifts  the flat mode from zero energy to 
$\varepsilon_{1 {\bf k}}\!\approx\! JS\sqrt{6d_M}$, where $d_M\!=\!\sqrt{3}D_z/J$,
see  Appendix~\ref{app0}.
The flat mode is also not entirely flat. This was interpreted \cite{Yildirim06} as a sign of a  
phenomenological next-nearest-neighbor superexchange $J_2$, ignoring 
its possible quantum origin. \cite{Chubukov92,ChZh14}
Since in the following we do not attempt a fully self-consistent calculation, the same interpretation suffices,
with an explicit expression for the dispersive flat mode given in Appendix~\ref{app0}.
Aside from this slight dichotomy with the origin 
of the flat-mode dispersion, the linear spin-wave theory seems to provide a spectacular 
account of the data without the need of any quantum effects.

However, we point out that the spectral weight is conspicuously missing from  experimental data
in the range of energies $15-19$meV in Fig.~\ref{Fig:wk_J2}(a), i.e. no signal has been detected there. 
While this feature has not been emphasized in Ref.~\onlinecite{Matan06} and one may argue that the
collected experimental data points were simply too sparse, the missing band  is strongly implied by our 
discussion, as it is exactly in the range of twice the energy of 
the flat mode, $2\varepsilon_{1{\bf k}}$, see Fig.~\ref{Fig:wk_J2}(a). 

\begin{figure*}
\includegraphics[width=1.85\columnwidth]{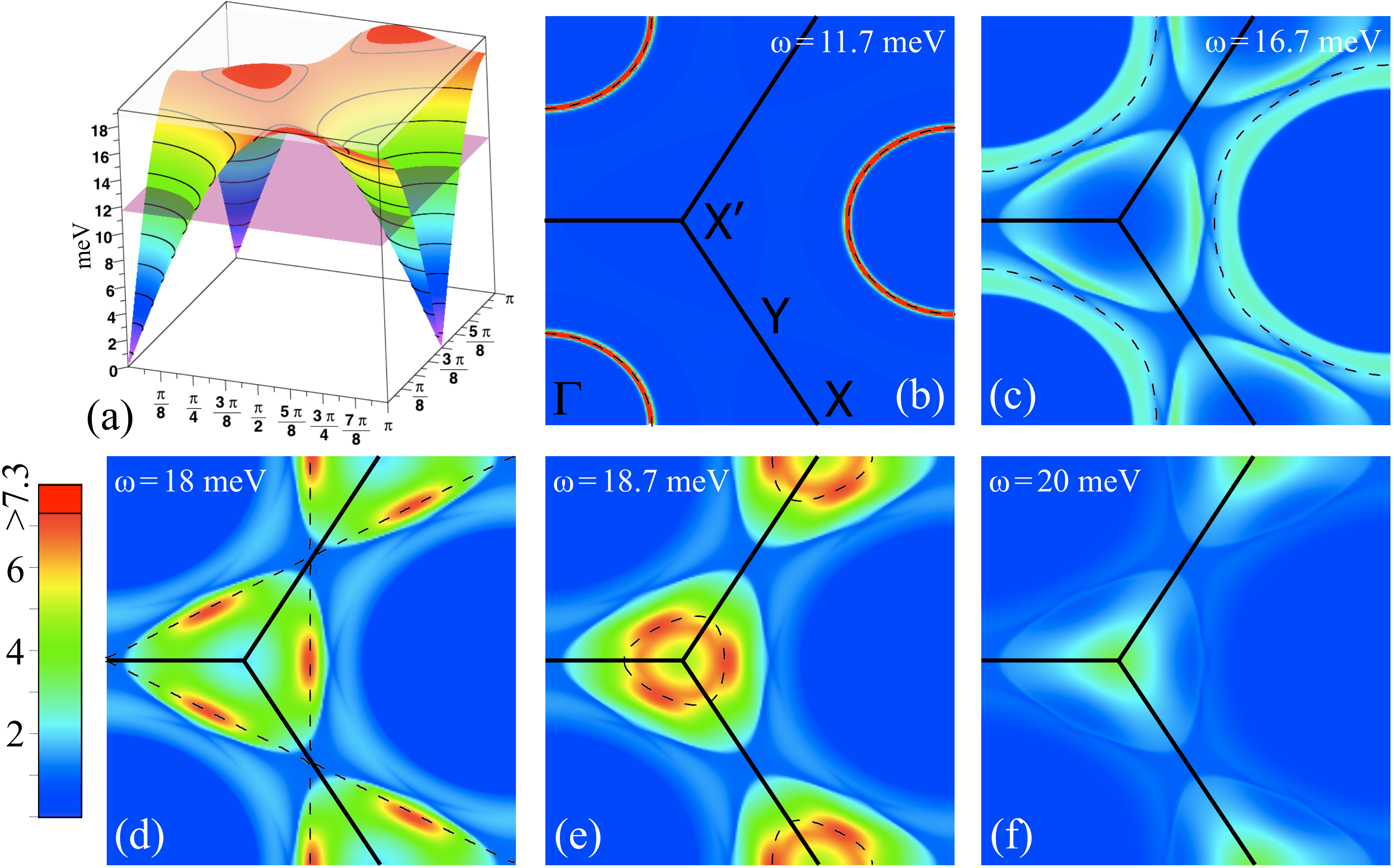}
\caption{(Color online)\ \ 
(a) A 3D plot of the magnon dispersion for Fe-jarosite within the linear spin-wave theory with  
planes of the  cuts in (b) and  (e). (b)-(f)
Intensity maps of $A_{\bf q}(\omega)$ in units of $(2SJ)^{-1}$ vs {\bf q} throughout the Brillouin zone for a set of  
energies. Intensity scale is as described in text.
Dashed black lines are peak positions  from the linear spin-wave theory. 
}
\label{fig:2DoS}
\end{figure*}

In Fig.~\ref{Fig:wk_J2}(b) we present the results of the on-shell calculation of 
$\Gamma_{{\bf k}}$ for the gapless dispersive mode using (\ref{Gamma})
with the flat-mode dispersion induced by $J_2$
for the same parameters as in  Fig.~\ref{Fig:wk_J2}(a), see Appendix~\ref{app0} for details.
As we discuss later, the dynamical structure factor allows to view modes selectively in 
different parts of the ${\bf k}$-space and in different polarizations.\cite{ChZh15} 
The results for the damping are combined with the energy $\varepsilon_{\bf k}$ of the mode 
with the shaded area representing half-width boundaries of a lorentzian peak,
$\varepsilon_{\bf k}\pm\Gamma_{{\bf k}}$. 
We have  also verified  \cite{ChZh15} that the effect of renormalization on the real part of the spectrum is minor, 
in agreement with  approximation in (\ref{Gamma}).

Our Fig.~\ref{Fig:wk_J2}(b)  demonstrates that the spin-wave excitation is well-defined
until a sharp threshold at about $2\varepsilon^{\rm min}_{1{\bf k}}$. 
Above that energy, the broadening  reaches about one third of the bandwidth 
signifying an overdamped spectrum, consistent with the missing spectral weight  in the experimental data.
The sharp transition implies a  threshold singularity and other spectral features that
are characteristic to the quasiparticle breakdown phenomenon
in quantum Bose liquids and $S\!=\!1/2$ spin-liquids. \cite{Zaliznyak,Zheludev}
There is a partial reconstruction of the spectrum at the energies above $2\varepsilon^{\rm max}_{1{\bf k}}$
where decays are no more resonant-like as indicated in the figure, i.e., occurring due to other, non-resonant channels,  
but  still providing a sizable broadening to the spectrum.

The non-resonant decays result in a typical broadening  $\Gamma\!\sim\!0.25J$, in accord with  similar 
results for the  triangular-lattice  \cite{triPRB09,tri06} and other frustrated spin systems. \cite{RMP13}
By contrast, the  broadening in the resonant-decay region in Fig.~\ref{Fig:wk_J2}(b) 
reaches $\Gamma\!\approx\! 1.7J$, an effect larger by a factor exceeding $2S$ for the 
considered $S\!=\!5/2$ model of Fe-jarosite. 
This is in a remarkable agreement with our qualitative discussion on the scaling of the resonance-like 
decay rate with $S$, provided after Eq.~(\ref{Gamma}) above.

We note that the broadening on the top of the band in the non-resonant region translates to less than 1meV, below
the experimental resolution of  Ref.~\onlinecite{Matan06}  in which  all the data were described as resolution-limited.
The current resolution of the neutron-scattering experiments 
is easily an order of magnitude higher. We also point out that our consideration is aimed at the 
strong qualitative features of the spectrum of a representative flat-band frustrated spin system, 
not on the minor quantitative details. As such, small discrepancies with some of the data may occur
due to, e.g., neglect of the in-plane DM terms, but should be considered as secondary.

\subsection*{Dynamical structure factor}

To demonstrate the effect of  decays, we performed a   calculation of the magnon spectral functions, 
$A_{\nu{\bf q}}(\omega)$, quantities directly related to the spin-spin dynamical correlation 
function via
\begin{eqnarray}
{\cal S}^{\alpha\alpha}({\bf q},\omega) \propto \int dt\,e^{i\omega t}\langle S^\alpha_{\bf q}(t)S^\alpha_{-\bf q}\rangle 
\propto \sum_{\nu} F^{\alpha}_{\nu\bf q} A_{\nu{\bf q}}(\omega) \,.
\label{Sqw}
\end{eqnarray}
Here, the kinematic formfactors $F^{\alpha}_{\nu\bf q}$ allow to ``filter out'' spectral
contributions of some of the modes 
to the in-plane and the out-of plane components of ${\cal S}({\bf q},\omega)$ in the portions of 
the ${\bf q}$-space  while highlighting the other ones: a phenomenon akin to the extinction of the Bragg peaks 
in the non-Bravias lattices. \cite{ChZh15} Using this feature,  we concentrate only on
one of the dispersive modes. 

A  dramatic view on the drastic transformations of the spectrum can be observed
in constant-energy cuts of the dynamical structure factor   in the range 
of energies affected by the resonance-like decays. In Fig.~\ref{fig:2DoS}, we present   intensity maps of 
such constant-energy cuts for  $A_{\nu{\bf q}}(\omega) $, a close proxy of the dynamical structure factor 
${\cal S}({\bf q},\omega)$, 
for  the dispersive magnon mode  for the energies ranging from 11.7meV to 20meV. 
The upper cut-off of the spectral function is chosen to correspond to the maximal height of the peaks 
in the non-resonant decay region in Fig.~\ref{Fig:wk_J2}(b) and  
translates into the broadening $\Gamma_{{\bf k}}\!\approx\!0.73$meV  for the Fe-jarosite values of $S$ and $J$, 
which should be resolvable by the modern neutron-scattering measurements.

The first of the cuts is below the threshold energy $2\varepsilon^{\rm min}_{1{\bf k}}$ and shows a very close
accord of the sharp-intensity peaks in  $A_{\nu{\bf q}}(\omega) $ with the expectations from the linear, non-interacting
spin-wave theory, shown by the dashed lines. The three subsequent  cuts,  Fig.~\ref{fig:2DoS}(c)-(e), are from 
within the resonant-decay band,   $2\varepsilon^{\rm min}_{1{\bf k}}\!<\!\omega\!<\!2\varepsilon^{\rm max}_{1{\bf k}}$,
where one can observe strong deviation from such expectations, massive redistribution of the spectral weight into   
different regions of the ${\bf q}$-space, and a multitude of intriguing ``shadow'' features, reflecting
van Hove singularities in the two-particle density of states of the decay products. \cite{triPRB09}
The last cut, Fig.~\ref{fig:2DoS}(f),
is, nominally, above the top of the magnon band and should be expected to show zero intensity everywhere. Instead,
it is also affected by the spectral weight redistribution and retains some of the features of the other cuts.
Altogether, Fig.~\ref{Fig:wk_J2}(b) and Fig.~\ref{fig:2DoS} offer a comprehensive theoretical insight into the 
non-trivial features of the dynamical structure factor of a flat-band kagome-lattice antiferromagnet, 
which originate from the  decays of magnetic excitations facilitated by the nonlinear couplings.

\section{Summary}
\label{Sec:summ} 
 
 To summarize, we have outlined a general scenario for drastic transformations in the spectra of frustrated magnets
that feature flat modes and have substantiated it by a  consideration of the spin-spin structure factor 
of the large-$S$ kagome-lattice system Fe-jarosite. Our study calls for further studies in these 
systems.

We  would also like to comment that  recently,
the  broad features in the spectra of magnetic systems have become a direct sign of 
fractionalized  excitations of prospective 
 spin-liquid phases. \cite{Coldea,Gardner,Balicas,Ronnow}
In this work, we have provided a case study of an excitation spectrum of a strongly frustrated but 
almost classical and well-ordered kagome-lattice antiferromagnet, for which we have demonstrated  
extremely strong broadening and even  a complete and spectacular wipe-out of a part of its spectrum. 
Here, the broad features are  due to flat or weakly dispersive modes, a hallmark feature of a 
variety of frustrated spin systems, and due to a non-collinearity of  spins in the ground state, 
again an outcome of  competing  interactions.
Thus,  this is also a cautionary tale, because the same reasons that may lead to 
the spin-liquid behavior may also favor strong coupling and decays among quasiparticles.

\begin{acknowledgments}

We acknowledge numerous enlightening discussions with Michael Zhitomirsky and a
useful conversation with Collin Broholm. 
We are thankful to Kittiwit Matan for sharing his previously published data. 
This work was supported by the U.S. Department of Energy,
Office of Science, Basic Energy Sciences under Award \# DE-FG02-04ER46174.
We would like to thank Aspen Center for Physics, where part of this work
was done, for hospitality. The work at Aspen was supported in part by NSF Grant No. PHYS-1066293.

\end{acknowledgments}

\appendix

\section{Technical details}
\label{app0}
\subsection{Spin-wave theory}

Following the approach outlined in Refs.~\onlinecite{Harris92,ChZh14}, 
one can diagonalize the harmonic part of the Hamiltonian in (\ref{HDM})
to obtain  the spin-wave energies\
\begin{equation}
\label{w1DM}
\varepsilon_{1\bf k}  = 2JS \sqrt{3d_M\left(1+d_M\right)/2} \, ,
\end{equation}
for  the ``flat mode,''  and 
\begin{eqnarray}
\label{w23DM}
&&\varepsilon_{2(3)\bf k}= 2JS \sqrt{1+d_M}\\
&&\phantom{\varepsilon_{2(3),\bf k}}\times\sqrt{1+d_M-\gamma_{\bf k} -d_M
\bigl(1 \pm \sqrt{1 + 8\gamma_{\bf k}}\,\bigr)/4}\, ,\nonumber \ \ \ \ 
\end{eqnarray}
for the  dispersive modes,  $d_M\!=\!\sqrt{3}D_z/J$ here. 

Corrections due to effective $J_2$ interactions can be taken into account perturbatively to yield\cite{ChZh15}
 the dispersion of the  ``flat mode''
\begin{eqnarray}
\label{w1J1J2Dz}
\varepsilon_{1\bf k}  &=& 2JS \sqrt{\left(3(1+d_M)/2+j_2
\left(1-\lambda^{(1)}_{1,{\bf k}}/2\right)\right)} \ \ \ \\
&&\times\sqrt{\left(d_M+j_2\left(1+\lambda^{(1)}_{1,{\bf k}}\right)\right)} + {\cal O}(j_2^{2})\, ,\nonumber
\end{eqnarray}
where $j_2=J_2/J$ and
\begin{eqnarray}
&&\lambda^{(1)}_{1,{\bf k}}=\left( f_2({\bf k})-f_1({\bf k})\right)/\left(1-\gamma_{\bf k}\right)  \, ,
\label{lambdaprime} \\
&&\mbox{with}  \ \ \ 
f_1({\bf k})=c_1^\prime c_1+c_2^\prime c_2+c_3^\prime c_3\,,   \nonumber\\
&&\phantom{\mbox{where}  \ \ \ }
f_2({\bf k})=c_1^\prime c_2 c_3+c_2^\prime c_1 c_3 + c_3^\prime c_1 c_2  \, ,\nonumber
\end{eqnarray}
with the shorthand notations $c_n = \cos(q_n)$, 
$c^\prime_1 \!=\!\cos(q_3\!+\!q_2)$, $c^\prime_2 \!=\!\cos(q_3\!-\!q_1)$, 
$c^\prime_3\!=\!\cos(q_1\!+\!q_2)$, where
$q_n\!=\!{\bf k}\cdot\bm{\delta}_n/2$,
and $\bm{\delta}_n$ are the primitive vectors of the kagome lattice.

The  diagonalization of the harmonic part of (\ref{HDM})
implies a two-step  procedure \cite{Harris92,ChZh14} with the unitary transformation 
of the original Holstein-Primakoff
bosons 
\begin{equation}
a_{\alpha\bf k} = \sum_\nu w_{\nu,\alpha}({\bf k})\, d_{\nu\bf k}\, ,
\label{linearT}
\end{equation}
followed by the usual Bogolyubov transformation for each of the individual species of $d$-boson, see 
Ref.~\onlinecite{ChZh14} for details and for the explicit form of the 
eigenvectors ${\bf w}_{\nu}=\left(w_{\nu,1}({\bf k}),w_{\nu,2}({\bf k}),w_{\nu,3}({\bf k})\right)$.

\subsection{Cubic terms}

Due to noncollinear 120$^\circ$ spin structure, 
cubic anharmonic coupling of the spin waves occurs. \cite{ChZh14,RMP13}  It originates from the 
$S_i^x S_j^z$ terms in  (\ref{HDM}), written in the local reference frame.  \cite{ChZh14}
In the bosonic representation they yield  
\begin{equation}
\hat{\cal H}_3 = J(1+d_M/3)\sqrt{\frac{S}{2}} \sum_{i,j} \sin\theta_{ij} \bigl( a_i^\dagger a_j^\dagger a_j +
\textrm{h.c.}\bigr)\, ,
\label{H3s}
\end{equation}
where $\theta_{ij}= \pm 120^\circ$ is the angle between two neighboring spins.

Assuming the spins in the ${\bf q}\!=\!0$   state, Fig.~\ref{Fig:DM}(a), and using the unitary and Bogolyubov 
transformations mentioned above gives the ``source'', $b^\dag b^\dag b^\dag$,
and the ``decay'', $b^\dag b^\dag b$, terms, see Ref.~\onlinecite{ChZh14} where the effects of the former 
 were discussed. 
The decay part of the Hamiltonian is 
\begin{equation}
\hat{\cal H}_3  = \frac{1}{2!} \frac{1}{\sqrt{N}} \sum_{\bf k+q=p}
\Phi^{\nu\mu\eta}_{\bf qk;p}\,
b_{\nu\bf q}^\dagger b_{\mu\bf k}^\dagger  b_{\eta\bf p} + \textrm{h.c.},
\label{Hd}
\end{equation}
with the  vertex
\begin{equation}
\Phi^{\nu\mu\eta}_{\bf qk;p}= -J\sqrt{\frac{3S}{2}}\;
\widetilde{\Phi}^{\nu\mu\eta}_{\bf qk;p} \,,
\label{V3d}
\end{equation}
which is explicitly $\propto\sqrt{S}$.
The symmetrized dimensionless vertex $\widetilde{\Phi}^{\nu\mu\eta}_{\bf qk;p}$ is given by
\begin{eqnarray}
&&\widetilde{\Phi}^{\nu\mu\eta}_{\bf qk;p}  =
F^{\nu\mu\eta}_{\bf qk;p} (u_{\nu\bf q}+v_{\nu\bf q})(u_{\mu\bf k}u_{\eta\bf p}+v_{\mu\bf k}v_{\eta\bf p})
\nonumber\\
&& \phantom{\widetilde{V}^{(d)\nu\mu\eta}_{\bf qk;p}} +
F^{\mu\eta\nu}_{\bf kpq} (u_{\mu\bf k}+v_{\mu\bf k})(u_{\nu\bf p}u_{\eta\bf q}+v_{\nu\bf p}v_{\eta\bf q}) \quad\quad
\label{V3d1}\\
&& \phantom{\widetilde{V}^{(d)\nu\mu\eta}_{\bf qk;p}} +
F^{\eta\nu\mu}_{\bf pqk} (u_{\eta\bf p}+v_{\eta\bf p})(u_{\nu\bf q}v_{\mu\bf k}+v_{\nu\bf q}u_{\mu\bf k}) ,
\nonumber
\end{eqnarray}
where $u_{\nu\bf k}$ and $v_{\nu\bf k}$ are the Bogolyubov parameters and 
the amplitudes $F^{\nu\mu\eta}_{\bf qkp}$ are given by
\begin{eqnarray}
F^{\nu\mu\eta}_{\bf qkp}=  \sum_{\alpha\beta}
\epsilon^{\alpha\beta\gamma}\cos(q_{\beta\alpha}) \,
w_{\nu,\alpha}({\bf q}) w_{\mu,\beta}({\bf k}) w_{\eta,\beta}({\bf p}),\ \
\label{F0}
\end{eqnarray}
where $\epsilon^{\alpha\beta\gamma}$ is the Levi-Civita
antisymmetric tensor,  and shorthand notations are
$q_{\beta\alpha}={\bf q}\bm{\rho}_{\beta\alpha}$
 and $\bm{\rho}_{\beta\alpha} = \bm{\rho}_\beta - \bm{\rho}_\alpha$, here $\bm{\rho}_\alpha$
 are the atom's positions within the unit cell.

\subsection{Self-energy, spectral function, and structure factor}

Using the standard diagrammatic rules for (\ref{Heff}), we obtain  
the second-order decay self-energy
\begin{eqnarray}
\Sigma_{\mu{\bf k}}(\omega)&=&\frac{1}{2}\sum_{{\bf q},\nu\eta}
\frac{|\Phi^{\nu\eta\mu}_{{\bf q},{\bf k}-{\bf q};{\bf k}}|^2}
{\omega - \varepsilon_{\nu\bf q} - \varepsilon_{\eta{\bf k}-{\bf q}}+i\delta},
\label{Sigma}
\end{eqnarray}
which contributes to the   $1/S$-correction to the magnon energy.
The magnon Green's function for the branch $\mu$ is given by (\ref{GF}).
Since  only the decay  terms  are responsible for the resonance-like decay
phenomena, one can  approximate the self-energy by its 
on-shell imaginary part, i.e. 
\begin{eqnarray}
\Sigma_{\mu{\bf k}}(\omega)\approx i {\rm Im}\Sigma_{\mu{\bf k}}(\varepsilon_{\mu{\bf k}})=-i\Gamma_{\mu{\bf k}}\,,
\label{Gamma2s}
\end{eqnarray}
which is given by (\ref{Gamma}).
Clearly, the dispersion of the flat mode in (\ref{w1J1J2Dz}) is crucial for the decays into two of them, as
otherwise this channel would produce essential singularity in (\ref{Sigma}) and in $\Gamma_{\mu{\bf k}}$.
With that, evaluation of the spectral function 
$A_{\mu{\bf k}}(\omega)=-(1/\pi){\rm Im}G_{\mu{\bf k}}(\omega)$
can be performed numerically. 

The diagonal components of the dynamical structure factor, or the spin-spin dynamical correlation function,
which contribute directly to the inelastic neutron-scattering cross section, are given by
\begin{eqnarray}
{\cal S}^{\alpha_0\alpha_0}({\bf q},\omega) = \int_{-\infty}^{\infty} \frac{dt}{2\pi}\,e^{i\omega t}\langle 
S^{\alpha_0}_{\bf q}(t)S^{\alpha_0}_{-\bf q}\rangle  \,,
\label{Sqws}
\end{eqnarray}
where $\alpha_0$ refers to the laboratory frame $\{x_0,y_0,z_0\}$.
Given the co-planar spin configuration, it is convenient to separate 
the in-plane and out-of-plane components of ${\cal S}^{\rm tot}({\bf q},\omega)$. 
Assuming equal contribution of all three $\alpha_0$  components to the cross section,
using the spin-wave mapping of spins on bosons with the two-step transformation described above,
after some algebra, one can obtain the leading contributions to the structure factor as
directly related to the spectral function
\begin{eqnarray}
{\cal S}^{\rm in(out)}({\bf q},\omega) =\sum_{\nu} F^{\rm in(out)}_{\nu\bf q} A_{\nu{\bf q}}(\omega) \,,
\label{Sqws1}
\end{eqnarray}
where $F^{\rm in(out)}_{\nu\bf q}$ are the kinematic formfactors.
It is important to note that the kinematic formfactors are modulated in the ${\bf q}$-space  and are suppressed 
in one of the Brillouin zones while are maximal in the others.\cite{ChZh15} 
This effect is characteristic to the non-Bravias lattices and 
is similar to the effect of extinction of some of the Bragg peaks in them. Because of that, one may be able to 
highlight spectral contribution of one of the magnon branches while 
``filtering out'' the others by selecting a particular component of the structure factor in a particular Brillouin zone.  
Our analysis demonstrates that 
the out-of plane component of ${\cal S}({\bf q},\omega)$ should be totally dominated by
 only one of the dispersive modes (gapless) in one of the three distinct 
Brillouin zones. This feature can be useful for the future neutron-scattering experiments.



\begin{thebibliography}{99}

\bibitem{Wannier}  G. H. Wannier, Phys. Rev. {\bf 79}, 357 (1950).

\bibitem{Villain}   A. Yoshimori, J. Phys. Soc. Jpn. {\bf 14}, 807 (1959).

\bibitem{Anderson}  P. W. Anderson, Science {\bf 235}, 1196 (1987).

\bibitem{Villain80}  J. Villain, R. Bidaux, J.-P. Carton, and R. Conte, Journal de Physique {\bf 41}, 1263 (1980).

\bibitem{Balents10}   L. Balents, Nature (London) {\bf 464}, 199 (2010).

\bibitem{Chalker92}  J. T. Chalker, P. C. W. Holdsworth, and E. F. Shender, Phys. Rev. Lett. {\bf 68}, 855 (1992).

\bibitem{Huse92}  D. A. Huse and A. D. Rutenberg, Phys. Rev. B {\bf 45}, 7536 (1992).

\bibitem{Harris92}  A. B. Harris, C. Kallin, and A. J. Berlinsky, Phys. Rev. B {\bf 45}, 2899 (1992).

\bibitem{Taillefumier14}  
M. Taillefumier, J. Robert, C. L. Henley, R. Moessner, and B. Canals, Phys. Rev. B {\bf 90}, 064419 (2014).

\bibitem{Reimers93}  J. N. Reimers and A. J. Berlinsky, Phys. Rev. B {\bf 48}, 9539 (1993).

\bibitem{Chubukov92} A. Chubukov, Phys. Rev. Lett. {\bf 69}, 832 (1992).

\bibitem{ChZh14}  A. L. Chernyshev and M. E. Zhitomirsky, Phys. Rev. Lett. {\bf 113}, 237202 (2014).

\bibitem{Jackeli15} G. Jackeli and A. Avella, arXiv:1504.01435 [cond-Mat] (2015).

\bibitem{Matan14} K. Matan, Y. Nambu, Y. Zhao, T. J. Sato, Y. Fukumoto, 
T. Ono, H. Tanaka, C. Broholm, A. Podlesnyak, and G. Ehlers, Phys. Rev. B {\bf 89}, 024414 (2014).

\bibitem{Matan06}   K. Matan, D. Grohol, D. G. Nocera, T. Yildirim, A. B. Harris, S. H. Lee, S. E. Nagler, 
and Y. S. Lee, Phys. Rev. Lett. {\bf 96}, 247201 (2006).

\bibitem{Yildirim06}  T. Yildirim and A. B. Harris, Phys. Rev. B {\bf 73}, 214446 (2006).

\bibitem{Petrenko}  
N. d’ Ambrumenil, O. A. Petrenko, H. Mutka, and P. P. Deen, arXiv:1501.03493 [cond-Mat] (2015).

\bibitem{Helton07}  J. S. Helton, K. Matan, M. P. Shores, E. A. Nytko, B. M. Bartlett, Y. Yoshida, 
Y. Takano, A. Suslov, Y. Qiu, J.-H. Chung, D. G. Nocera, and Y. S. Lee, Phys. Rev. Lett. {\bf 98}, 107204 (2007).

\bibitem{Lee12} 
T.-H. Han, J. S. Helton, S. Chu, D. G. Nocera, J. A. Rodriguez-Rivera, C. Broholm, and Y. S. Lee, 
Nature (London) {\bf 492}, 406 (2012).

\bibitem{Matan10}  
K. Matan, T. Ono, Y. Fukumoto, T. J. Sato, J. Yamaura, M. Yano, K. Morita, and H. Tanaka, Nat. Phys. {\bf 6}, 865 (2010).

\bibitem{Yan11}  S. Yan, D. A. Huse, and S. R. White, Science {\bf 332}, 1173 (2011).

\bibitem{Mambrini00}  M. Mambrini and F. Mila, Eur. Phys. J. B {\bf 17}, 651 (2000).

\bibitem{Iqbal13}  Y. Iqbal, F. Becca, S. Sorella, and D. Poilblanc, Phys. Rev. B {\bf 87}, 060405 (2013).

\bibitem{Rousochatzakis14}  I. Rousochatzakis, Y. Wan, O. Tchernyshyov, and F. Mila, Phys. Rev. B {\bf 90}, 100406 (2014).

\bibitem{Elhajal02}  M. Elhajal, B. Canals, and C. Lacroix, Phys. Rev. B {\bf 66}, 014422 (2002).

\bibitem{Zorko13}  A. Zorko, F. Bert, A. Ozarowski, J. van Tol, D. Boldrin, A. S. Wills, and P. Mendels, 
Phys. Rev. B {\bf 88}, 144419 (2013).

\bibitem{Yoshida12}  H. Yoshida, Y. Michiue, E. Takayama-Muromachi, and M. Isobe, J. Mater. Chem. {\bf 22}, 18793 (2012).

\bibitem{RMP13}  M. E. Zhitomirsky and A. L. Chernyshev, Rev. Mod. Phys. {\bf 85}, 219 (2013).

\bibitem{Henley93}  J. von Delft and C. L. Henley, Phys. Rev. B {\bf 48}, 965 (1993).

\bibitem{Gotze15}  O. G\"{o}tze and J. Richter, Phys. Rev. B {\bf 91}, 104402 (2015).

\bibitem{Zaliznyak} M. B. Stone, I. A. Zaliznyak, T. Hong, C. L. Broholm, and D. H. Reich,   Nature (London)  {\bf 440}, 187 (2006).

\bibitem{Zheludev}   T. Masuda, A. Zheludev, H. Manaka, L.-P. Regnault, J.-H. Chung, and Y. Qiu, Phys. Rev. Lett. 
{\bf 96}, 047210 (2006).


\bibitem{Balents_honeycomb}  C. Wu, D. Bergman, L. Balents, and S. Das Sarma, Phys. Rev. Lett. {\bf 99}, 070401 (2007).

\bibitem{tri06}  A. L. Chernyshev and M. E. Zhitomirsky, Phys. Rev. Lett. {\bf 97}, 207202 (2006).

\bibitem{Zh06} M. E. Zhitomirsky, Phys. Rev. B {\bf 73}, 100404 (2006).

\bibitem{Vojta} L. Fritz, R. L. Doretto, S. Wessel, S. Wenzel, S. Burdin, and M. Vojta, Phys. Rev. B {\bf 83}, 174416 (2011).

\bibitem{Cepas08}  O. C\'{e}pas, C. M. Fong, P. W. Leung, and C. Lhuillier, Phys. Rev. B  {\bf 78}, 140405 (2008).

\bibitem{Zorko08}  A. Zorko, S. Nellutla, J. van Tol, L. C. Brunel, F. Bert, F. Duc, J.-C. Trombe, 
M. A. de Vries, A. Harrison, and P. Mendels, Phys. Rev. Lett. {\bf 101}, 026405 (2008).

\bibitem{francisite15} 
I. Rousochatzakis, J. Richter, R. Zinke, and A. A. Tsirlin, Phys. Rev. B {\bf 91}, 024416 (2015).

\bibitem{ChZh15}  A. L. Chernyshev and M. E. Zhitomirsky, to be publushed.

\bibitem{triPRB09}  A. L. Chernyshev and M. E. Zhitomirsky, Phys. Rev. B  {\bf 79}, 144416 (2009).

\bibitem{Ronnow}  B. Dalla Piazza, M. Mourigal, N. B. Christensen, G. J. Nilsen, P. Tregenna-Piggott, 
T. G. Perring, M. Enderle, D. F. McMorrow, D. A. Ivanov, and H. M. R{\o}nnow, Nat. Phys.  {\bf 11}, 62 (2015).

\bibitem{Gardner} H. D. Zhou, C. Xu, A. M. Hallas, H. J. Silverstein, C. R. Wiebe, I. Umegaki,
 J. Q. Yan, T. P. Murphy, J.-H. Park, Y. Qiu, J. R. D. Copley, J. S. Gardner, and Y. Takano, Phys. Rev. Lett.  {\bf 109}, 267206 (2012).

\bibitem{Balicas} H. D. Zhou, E. S. Choi, G. Li, L. Balicas, C. R. Wiebe, Y. Qiu, J. R. D. Copley, and 
J. S. Gardner, Phys. Rev. Lett.  {\bf 106}, 147204 (2011).

\bibitem{Coldea}  R. Coldea, D. A. Tennant, A. M. Tsvelik, and Z. Tylczynski, Phys. Rev. Lett.  {\bf 86}, 1335 (2001).



\end{thebibliography}
\end{document}